\documentclass[12pt,a4paper]{article}
\usepackage[utf8]{inputenc}
\usepackage{float}

\usepackage{graphicx}
\usepackage[margin=25mm]{geometry}
\usepackage[font=sf]{caption}

\usepackage{amsmath, amsfonts, amssymb}
\usepackage{siunitx}
\usepackage{hyperref}
\usepackage{natbib}
\usepackage{booktabs}

\title{Efficient Electric Vehicle Charging Allocation:\\
A Two-Stage Optimization and Participation Analysis
\thanks{This manuscript is a highly preliminary draft.
Both the model and the simulation design will be extended in future versions.
For readability, we adopt the simplest formulations and descriptions that convey the main ideas.}}
\author{LIU Ruiwu \& ZHU Yangjian}
\date{}

\begin{document}
\maketitle

\vspace{0.5em}
\begin{abstract}
Electric vehicles (EVs) require substantially longer refueling times than gasoline vehicles, which can generate severe congestion at charging stations when demand concentrates. We propose a two-stage allocation framework for EV charging networks. In Stage~1, a central coordinator determines station-level admission quotas to control worst-station delay using a queue-informed congestion metric. In Stage~2, given these quotas and feasibility constraints (e.g., reachability), the coordinator solves a utility-maximizing capacitated assignment to allocate EVs across stations. To keep Stage~2 tractable while capturing heterogeneous charging needs, we precompute each EV--station pair's optimal charging amount in closed form under a battery-capacity constraint and then solve a transportation/assignment problem. Finally, we introduce a reduced-form participation model to characterize adoption thresholds under network benefits, spillovers, and coordination costs. Numerical experiments illustrate substantial reductions in worst-case congestion with limited impact on average utility, and highlight scaling patterns as the number of stations increases.
\end{abstract}

\begin{quotation}
\end{quotation}

\section{Introduction}\label{sec:intro}
EV adoption has increased rapidly in many regions, while charging infrastructure expansion often lags behind. Unlike gasoline refueling, EV charging can take tens of minutes or longer, so demand concentration can cause long queues and extreme waiting times at popular stations. Uncoordinated charging decisions also create congestion externalities: an EV choosing a station increases delays for others, while underutilized stations remain idle.

A large literature studies EV charging using queueing models, dynamic pricing, routing algorithms, and game-theoretic approaches (e.g., Stackelberg-type interactions and matching-based formulations). In addition, participation issues arise when coordination requires users to share information or comply with platform recommendations. Similar to coalition-formation problems in International Environmental Agreement (IEA) analysis, nonparticipants may still benefit from spillovers created by a coordinated subgroup, which generates free-riding incentives and adoption thresholds.

\paragraph{Contributions}
Our paper contributes to the EV charging allocation literature along several dimensions.
First, while much of the existing work studies either congestion using queueing models or allocation using pricing/routing/matching mechanisms, we provide a unified two-stage framework that explicitly separates \emph{system-level congestion control} from \emph{user-level welfare maximization}.
Stage~1 introduces station-level admission quotas that directly target tail congestion by minimizing the worst-station delay, thereby internalizing congestion externalities at the system level. This contrasts with approaches that focus on individual best responses (e.g., uncoordinated routing) or primarily improve average performance without explicitly controlling worst-case congestion.

Second, relative to stable-matching-based scheduling (e.g., Gale--Shapley-type mechanisms used in prior EV charging applications), our Stage~2 allocation is formulated as a welfare-maximizing capacitated assignment under quotas.
This reflects the practical reality that stations typically do not ``reject'' customers in the way matching models allow, and it yields a directly implementable allocation rule.
To keep the model tractable while incorporating heterogeneous charging needs, we precompute each EV--station pair's optimal charging amount in closed form under a battery-capacity constraint, reducing Stage~2 to a transportation/assignment problem.

Third, our formulation clarifies two economically meaningful ways of accounting for congestion in welfare.
When $\eta=0$, congestion is internalized indirectly through the quota constraints (a hard congestion-control layer);
when $\eta>0$, congestion is priced into welfare as an explicit time-cost term (a soft adjustment).
This provides a clean bridge between operational constraints and welfare interpretation that is often implicit in the existing literature.

Finally, beyond operational allocation, we study voluntary participation using a reduced-form adoption model motivated by network externalities and coalition stability.
Unlike most scheduling papers that assume full participation, our participation module characterizes adoption thresholds in the presence of spillovers (free-riding) and coordination costs, highlighting when a coordinated platform is sustainable and why ``cold-start'' incentives may be required.

\paragraph{Paper organization.}
Section~\ref{sec:model} introduces the two-stage model. Section~\ref{sec:numerical} presents numerical experiments and simulation results. Section~\ref{sec:participation} presents the participation model. Section~\ref{sec:discussion} discusses limitations and extensions, and Section~\ref{sec:conclusion} concludes.

\section{Model}\label{sec:model}

\subsection{Time, sets, and notation}
Time is discrete, indexed by $k=0,1,2,\dots$, with scheduling interval length $T>0$.
(For notational simplicity, one may normalize $T=1$; then rates such as $\mu_i$ are measured per interval.)

Let $\mathcal S=\{1,\dots,S\}$ be the set of charging stations (CSs), indexed by $i\in\mathcal S$.
In each interval $k$, a set of EVs $\mathcal N(k)$ arrives, with $N(k)=|\mathcal N(k)|$ and $\gamma(k)\equiv N(k)$ ($\gamma(k)$ denotes the exogenous inflow of new charging requests at epoch $k$). 

Station $i$ has: (i) $c_i\in\mathbb Z_{+}$ chargers; (ii) per-charger service rate $\mu_i>0$; (iii) unit price $p_i\ge 0$ (extensions allow $p_i(k)$).
Each EV $n\in\mathcal N(k)$ has $SoC_n(k)\in[0,1]$, distance $d_{ni}\ge 0$ to station $i$, and willingness-to-pay cap $p_n^{\max}\ge 0$.
Let $\kappa>0$ be a range conversion parameter (distance per unit $SoC$). The feasible station set for EV $n$ is
\begin{equation}\label{eq:feasible_set}
\mathcal S_n(k)=\{i\in\mathcal S:\ d_{ni}\le \kappa\cdot SoC_n(k)\}. 
\end{equation}
We assume $\mathcal S_n(k)\neq\varnothing$ for all $n$ (otherwise a fallback rule applies).
For example, $d_{ni}\le \kappa SoC_n(k)$ means EV $n$ can reach station $i$ if its remaining charge supports the travel distance.

\subsection{Stage 1: flow--queue coupling and congestion control}\label{sec:stage1}
\paragraph{State and flow variables.}
Let $x_i(k)\ge 0$ denote the number of EVs \emph{in station $i$} at the beginning of interval $k$ (waiting + in service).
Let $f_i(k)\ge 0$ be the number of EVs admitted/routed to station $i$ during interval $k$ (vehicles per interval),
and $g_i(k)\ge 0$ the number departing station $i$ during interval $k$ (vehicles per interval).
The station queue evolves as
\begin{equation}\label{eq:queue_dynamics}
x_i(k+1)=x_i(k)+f_i(k)-g_i(k),\qquad \forall i\in\mathcal S.
\end{equation}
Flow conservation is
\begin{equation}\label{eq:flow_conservation}
\sum_{i\in\mathcal S} f_i(k)=\gamma(k),\qquad f_i(k)\ge 0.
\end{equation}

\paragraph{Arrival-rate estimation via historical smoothing (moving average).}
We distinguish per-interval counts $f_i(k)$ from the continuous-time arrival rate $\lambda_i(k)$ used in queueing performance.
We estimate
\begin{equation}\label{eq:lambda_smoothing_ma}
\lambda_i(k)=\frac{1}{H}\sum_{\ell=0}^{H-1} f_i(k-\ell),\qquad H\in\mathbb Z_{+},\ \forall i\in\mathcal S.
\end{equation}
(Note that $f_i(k)$ enters $\lambda_i(k)$ through the moving-average estimator, so Stage~1 has a short-memory effect over $H$ epochs.)
For early intervals $k<H$, one may use a shorter window or initialize $f_i(k)=0$ for $k<0$.

\subsubsection{Queueing background (M/M/$c_i$) and a tractable output approximation}
Within each interval, station $i$ is approximated as an M/M/$c_i$ queue with arrival rate $\lambda_i(k)$ and per-server service rate $\mu_i$.
Define offered load and utilization:
\begin{equation}\label{eq:rho_def}
\rho_i(k)=\frac{\lambda_i(k)}{\mu_i},\qquad
\upsilon_i(k)=\frac{\lambda_i(k)}{c_i\mu_i}=\frac{\rho_i(k)}{c_i}.
\end{equation}
Stability requires
\begin{equation}\label{eq:stability}
\lambda_i(k)<c_i\mu_i,\qquad \forall i,k.
\end{equation}
Let $P_{0,i}(k)$ denote the probability that station $i$ is empty in steady state. Then
\begin{equation}\label{eq:P0_mm_c}
P_{0,i}(k)=
\left[
\sum_{n=0}^{c_i-1}\frac{\rho_i(k)^n}{n!}
+\frac{\rho_i(k)^{c_i}}{c_i!\left(1-\frac{\rho_i(k)}{c_i}\right)}
\right]^{-1}.
\end{equation}
The expected number of EVs \emph{in station $i$} (waiting + in service) admits the standard Erlang-C form
\begin{equation}\label{eq:L_mm_c}
\mathbb E[x_i\mid \lambda_i(k)]
=
\frac{\rho_i(k)^{c_i+1}}{c_i\,c_i!\left(1-\frac{\rho_i(k)}{c_i}\right)^2}P_{0,i}(k)+\rho_i(k).
\end{equation}

Standard Erlang-C formulas map $\lambda_i(k)$ to $\mathbb E[x_i\mid \lambda_i(k)]$ (via $P_{0,i}(k)$, etc.). (While the Erlang-C expressions provide a standard steady-state mapping from $\lambda_i(k)$ to expected congestion,
our Stage~1 implementation relies on the explicit dynamic closure \eqref{eq:g_hat}--\eqref{eq:departure_count}
to update $(x_i(k),W_i(k))$ epoch-by-epoch in simulation and optimization.) However, for dynamic updates \eqref{eq:queue_dynamics} we require an explicit outflow function $g_i$ as a function of the state $x_i(k)$. Inverting the steady-state mapping to obtain $g_i=g_i(x_i)$ in closed form is generally intractable for $c_i>1$. Following the tractable approximation approach in \citet{2017Distributed} (and related distributed queue-control work), we close the dynamics using an explicit saturating outflow map:
\begin{equation}\label{eq:g_hat}
\hat g_i(x)=c_i\mu_i\frac{x}{1+x},\qquad x\ge 0,
\end{equation}
so departures per interval are
\begin{equation}\label{eq:departure_count}
g_i(k)=\,\hat g_i(x_i(k)) = \,c_i\mu_i\frac{x_i(k)}{1+x_i(k)}.
\end{equation}

\paragraph{Delay metric (Little's law).}
Because $x_i(k)$ counts EVs in station $i$ (waiting + in service), Little's law implies that the expected \emph{time in system} (sojourn time) is
\begin{equation}\label{eq:little_sojourn}
W_i(k)=\frac{x_i(k)}{\lambda_i(k)}.
\end{equation}
(Optionally, queueing delay can be defined as $W_{q,i}(k)=W_i(k)-1/\mu_i$.)

\subsubsection{Stage 1 optimization and numerical solution}
Stage~1 chooses admissions $\{f_i(k)\}$ to reduce worst-station delay. Introducing an auxiliary variable $z$ yields the epigraph form (equivalent to ):
\begin{align}
\min_{\{f_i(k)\},\,z}\quad & z \label{eq:stage1_obj}\\
\text{s.t.}\quad &
W_i(k)\le z,\qquad \forall i\in\mathcal S, \label{eq:stage1_z}\\
&\sum_{i\in\mathcal S} f_i(k)=\gamma(k),\qquad f_i(k)\ge 0,\ \forall i, \label{eq:stage1_sum}\\
&\lambda_i(k)\ \text{is given by \eqref{eq:lambda_smoothing_ma}},\qquad \forall i, \label{eq:stage1_smooth}\\
&\lambda_i(k)<c_i\mu_i,\qquad \forall i. \label{eq:stage1_stab}
\end{align}
The state evolves by \eqref{eq:queue_dynamics}--\eqref{eq:departure_count}.
Problem \eqref{eq:stage1_obj}--\eqref{eq:stage1_stab} can be solved numerically at each epoch using a standard nonlinear solver.
Alternatively, one may biselect on $z$ and check feasibility; feasibility is monotone in $z$.

\subsection{Stage 2: utility-maximizing assignment under Stage-1 quotas}\label{sec:stage2}

\paragraph{Binary assignment decision.}
Let $x_{ni}(k)\in\{0,1\}$ indicate whether EV $n$ is assigned to station $i$ in interval $k$.
Feasibility requires $x_{ni}(k)=0$ for all $i\notin\mathcal S_n(k)$.
Each EV is assigned to exactly one feasible station:
\begin{equation}\label{eq:assign_one}
\sum_{i\in\mathcal S_n(k)} x_{ni}(k)=1,\qquad \forall n\in\mathcal N(k).
\end{equation}
Stage 1 admissions $f_i(k)$ act as station quotas:
\begin{equation}\label{eq:assign_quota}
\sum_{n\in\mathcal N(k)} x_{ni}(k)\le f_i(k),\qquad \forall i\in\mathcal S.
\end{equation}

\paragraph{Charging amount: precomputed optimal $E_{ni}^{\ast}(k)$.}
Let $E_{ni}(k)$ denote the energy fraction charged by EV $n$ if it uses station $i$ in interval $k$.
Battery-capacity constraint:
\begin{equation}\label{eq:E_constraint}
0\le E_{ni}(k)\le 1-SoC_n(k).
\end{equation}
We capture heterogeneity in charging preference via a curvature/anxiety parameter. Each EV has a baseline $\bar s_n>0$ and we allow state dependence:
\begin{equation}\label{eq:S_def}
S_n(k)=\bar s_n\big(1+\alpha(1-SoC_n(k))\big),\qquad \alpha\ge 0.
\end{equation}
Define the charging benefit function for pair $(n,i)$:
\begin{equation}\label{eq:phi_def}
\phi_{ni}(E)= -\frac12 S_n(k)E^2 + \big(p_n^{\max}-p_i\big)E.
\end{equation}
For fixed $(n,i)$, solve the one-dimensional concave problem
\begin{equation}\label{eq:E_star_problem}
E^{\ast}_{ni}(k)=\arg\max_{0\le E\le 1-SoC_n(k)} \phi_{ni}(E),
\end{equation}
which admits the closed-form solution:
\begin{equation}\label{eq:E_star_closed}
E^{\ast}_{ni}(k)=\min\left\{\max\left\{\frac{p_n^{\max}-p_i}{S_n(k)},\,0\right\},\,1-SoC_n(k)\right\}.
\end{equation}

\paragraph{Utility and the role of congestion.}
Let $\tau>0$ be the distance disutility weight.
Optionally, the Stage-1 congestion signal enters welfare via a time-cost weight $\eta\ge 0$.
For $i\in\mathcal S_n(k)$, define:
\begin{equation}\label{eq:utility_stage2}
u_{ni}(k)=\phi_{ni}\!\big(E^{\ast}_{ni}(k)\big)-\tau d_{ni}-\eta W_i(k).
\end{equation}
Setting $\eta=0$ omits an explicit delay term in Stage~2 (congestion is controlled indirectly via quotas), while $\eta>0$ yields a fully closed welfare interpretation.

\paragraph{Stage 2 optimization problem.}
The platform solves the capacitated assignment problem:
\begin{align}
\max_{\{x_{ni}(k)\}}\quad &
\sum_{n\in\mathcal N(k)}\sum_{i\in\mathcal S_n(k)} u_{ni}(k)\,x_{ni}(k)\label{eq:stage2_obj}\\
\text{s.t.}\quad &
\sum_{i\in\mathcal S_n(k)} x_{ni}(k)=1,\quad \forall n\in\mathcal N(k), \label{eq:stage2_one}\\
&\sum_{n\in\mathcal N(k)} x_{ni}(k)\le f_i(k),\quad \forall i\in\mathcal S, \label{eq:stage2_quota}\\
&x_{ni}(k)\in\{0,1\},\quad \forall n,i. \label{eq:stage2_bin}
\end{align}
This is a transportation/assignment problem, directly implementable and simulatable (often solvable efficiently via min-cost flow / LP methods).

\section{Numerical Experiments and Simulation Results}\label{sec:numerical}

We simulate the system in discrete epochs. In each epoch, new EV requests arrive, the platform determines station quotas (Stage~1), and then assigns EVs under quotas (Stage~2). For baseline comparison, we also consider a free-choice strategy in which EVs select the nearest feasible station, and a matching-based baseline (Yoshihara/Gale--Shapley) when applicable.

\subsection{Simulation design}
\paragraph{Time resolution and arrivals.}
We simulate the system over discrete decision epochs. At each epoch $k$, a batch of new EV charging requests arrives.
The number of new requests $\gamma(k)$ is generated by a Poisson distribution with parameter
$\Lambda$ (e.g., $\Lambda=30$). All arriving EVs are assumed to require charging.

\paragraph{Cold start and per-epoch sequence.}
Stage~1 uses the smoothed arrival-rate estimator (moving average) to compute $\lambda_i(k)$, which requires past inflow data.
Therefore, we adopt the following initialization:
(i) in the first epoch, quotas $\{f_i(1)\}$ are allocated proportionally to incoming demand (e.g., proportional to station capacity);
(ii) only the Stage~2 allocation is solved in the first epoch;
(iii) from the second epoch onward, we solve Stage~1 to determine quotas, and then solve Stage~2 under these quotas.

More precisely, for each epoch $k\ge 2$:
\begin{enumerate}
    \item Observe the station states $\{x_i(k)\}$ and new arrivals $\mathcal N(k)$ with total size $\gamma(k)$.
    \item \textbf{Stage 1 (quota control):} solve the min--max congestion problem to obtain station quotas $\{f_i(k)\}$.
    \item Update smoothed arrival-rate estimates $\lambda_i(k)$ and compute departures using the explicit saturating outflow approximation;
    update queue states via $x_i(k+1)=x_i(k)+f_i(k)-g_i(k)$.
    \item \textbf{Stage 2 (assignment):} compute each feasible EV--station pair's utility (including the closed-form optimal charging amount),
    and solve the capacitated assignment problem under quotas.
\end{enumerate}

\paragraph{Baselines.}
We compare our two-stage mechanism with:
\begin{itemize}
    \item \textbf{Free choice (nearest-station) baseline:} each EV selects the nearest feasible station.
    \item \textbf{Matching-based baseline (Yoshihara/Gale--Shapley):} the algorithm proposed in \citet{yoshihara_non-cooperative_2020},
    under the same environment configuration (when applicable).
\end{itemize}

\subsection{Parameter setting}
Tables~\ref{tab:ev_cs_param}--\ref{tab:ev_param} summarize the core parameters used for EVs and charging stations.

\begin{table}[H]
\centering
\begin{tabular}{c|c|p{8cm}}
\toprule
Class & Parameter & Description \\
\midrule
EV & $SoC_n(k)$ & State of charge (normalized to $[0,1]$) \\
EV & $p_n^{\max}$ & Willingness-to-pay cap \\
EV & $\bar s_n,\alpha$ & Curvature/anxiety parameters in \eqref{eq:S_def} \\
EV & position & Entry location (used to compute $d_{ni}$) \\
\midrule
CS & $c_i$ & Number of chargers \\
CS & $\mu_i$ & Service rate per charger \\
CS & $p_i$ & Price per unit charge \\
CS & location & Station location \\
\midrule
System & $\Lambda$ & Poisson arrival intensity for $\gamma(k)$ \\
System & $H$ & Moving-average window length for $\lambda_i(k)$ \\
\bottomrule
\end{tabular}
\caption{Parameter descriptions for EVs and charging stations (core).}
\label{tab:ev_cs_param}
\end{table}

\begin{table}[H]
\centering
\begin{tabular}{c|c|c|c}
\toprule
CS idx & $c_i$ & location & price \\
\midrule
1 & 2 & 8  & 62 \\
2 & 1 & 12 & 60 \\
3 & 3 & 20 & 58 \\
\bottomrule
\end{tabular}
\caption{Example station configuration.}
\label{tab:cs_param}
\end{table}

\begin{table}[H]
\centering
\begin{tabular}{c|p{9cm}}
\toprule
Parameter & Setting \\
\midrule
$p_n^{\max}$ & Randomly assigned (e.g., uniform on $[80,120]$) \\
Position & Randomly assigned over a line segment (e.g., $[0,20]$) \\
Initial $SoC_n$ & Randomly assigned (e.g., uniform on $[0.1,0.9]$) \\
Anxiety type & Randomly assigned via $(\bar s_n,\alpha)$ \\
\bottomrule
\end{tabular}
\caption{Example EV parameterization.}
\label{tab:ev_param}
\end{table}

\subsection{Results: free choice vs two-stage optimization}
Stage~1 quotas primarily target \emph{extreme congestion} by preventing demand from concentrating on already overloaded stations.
Stage~2 then allocates EVs within these quotas to maximize aggregate utility, so the welfare impact of congestion control is moderated.

\begin{figure}[htbp]
\centering
\begin{minipage}[b]{0.48\textwidth}
\centering
\includegraphics[width=\linewidth]{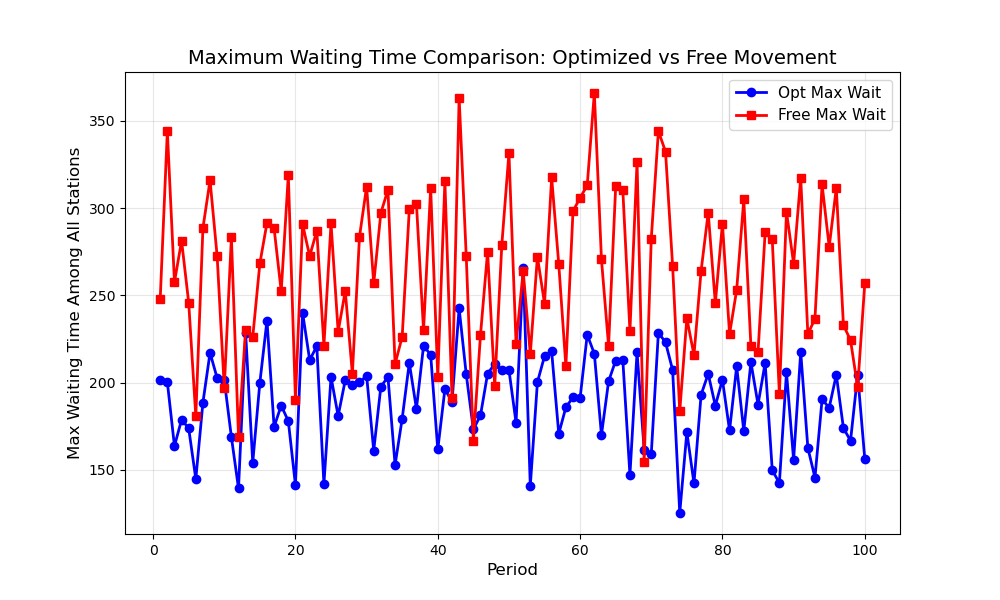}
\caption{Time saving: free vs optimization}
\label{fig:time}
\end{minipage}
\hfill
\begin{minipage}[b]{0.48\textwidth}
\centering
\includegraphics[width=\linewidth]{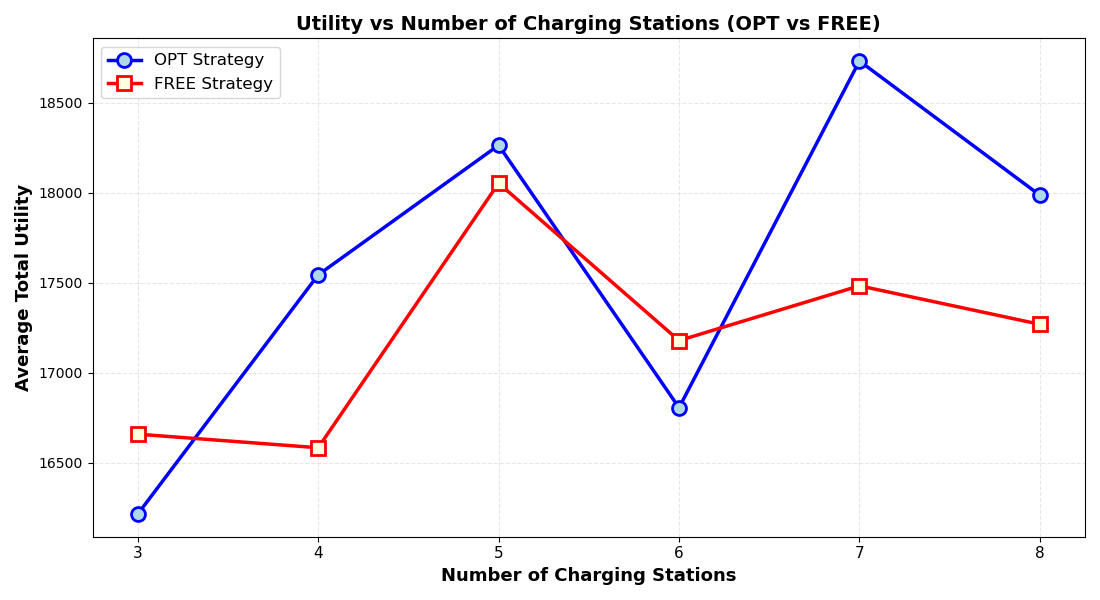}
\caption{Average utility vs number of stations}
\label{fig:avgutility}
\end{minipage}
\end{figure}

We observe that worst-case congestion measures (e.g., maximum queue length and time-related metrics)
are substantially improved under the two-stage mechanism.
At the same time, average utility can remain close to the free-choice baseline in small networks, because limited station options constrain how much utility can be improved purely through assignment.
As the number of stations increases, the feasible choice set expands and the utility-maximizing assignment becomes more effective,
so the gap in average utility can widen in favor of the optimized allocation (Figure~\ref{fig:avgutility}).
\begin{figure}[htbp]
\centering
\begin{minipage}[b]{0.48\textwidth}
\centering
\includegraphics[width=\linewidth]{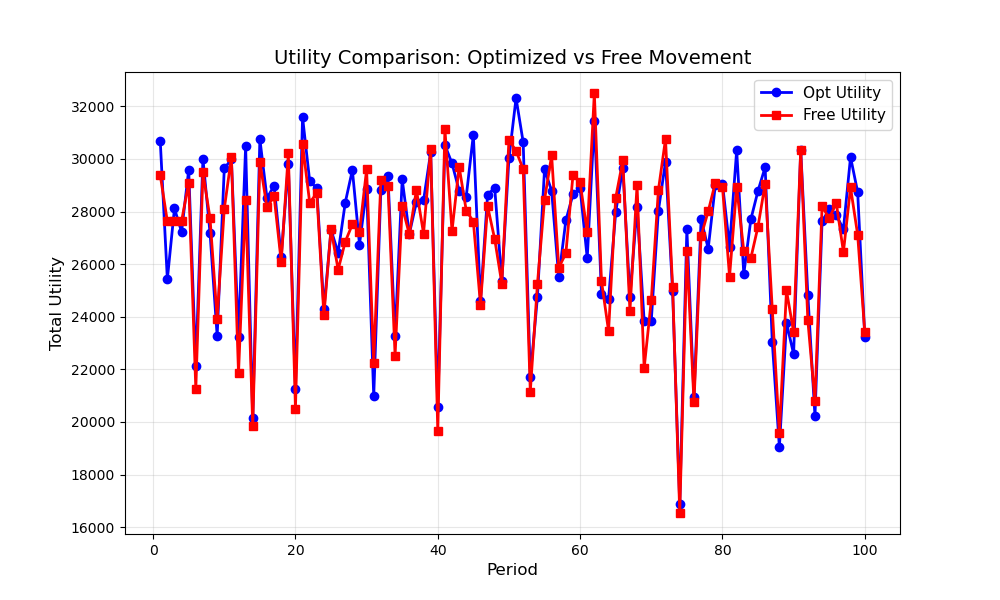}
\caption{Utility: free vs optimization}
\label{fig:utility}
\end{minipage}
\hfill
\begin{minipage}[b]{0.48\textwidth}
\centering
\includegraphics[width=\linewidth]{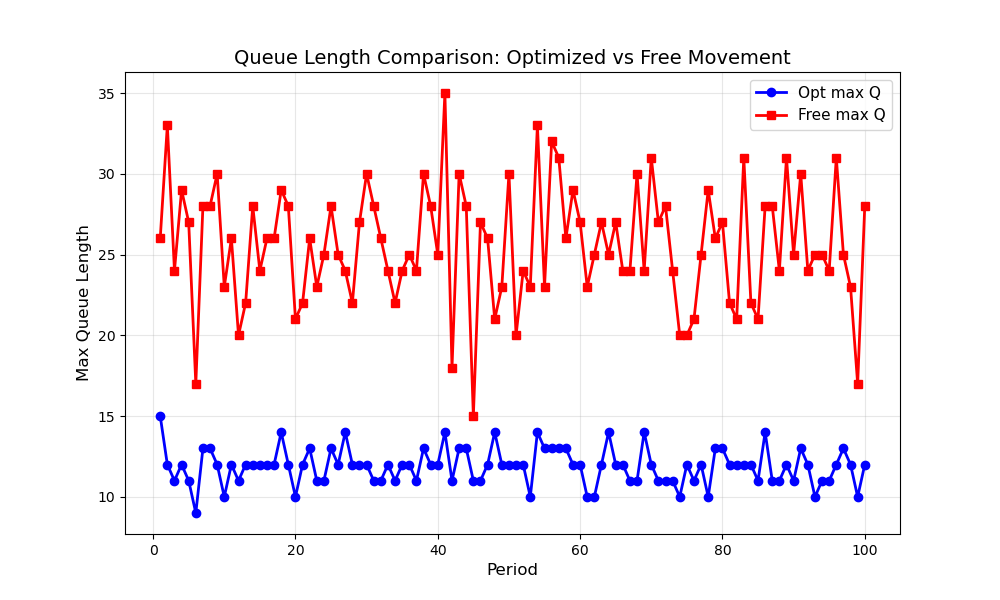}
\caption{Queue length: free vs optimization}
\label{fig:queue}
\end{minipage}
\end{figure}

\paragraph{Interpreting $\eta=0$ vs $\eta>0$ (under construction).}
When $\eta=0$, congestion is internalized indirectly through the quota constraints in \eqref{eq:assign_quota}, creating a \emph{hard} congestion-control layer. When $\eta>0$, congestion enters utility directly as a \emph{soft} welfare adjustment; marginal EVs may be shifted away from more congested stations, potentially further reducing delay at the cost of small utility trade-offs (depending on $\eta$). If desired, one can report sensitivity results by varying $\eta$ while keeping the same baseline settings.

\subsection{Comparison with matching-based baseline}
\begin{figure}[htbp]
\centering
\includegraphics[width=0.55\linewidth]{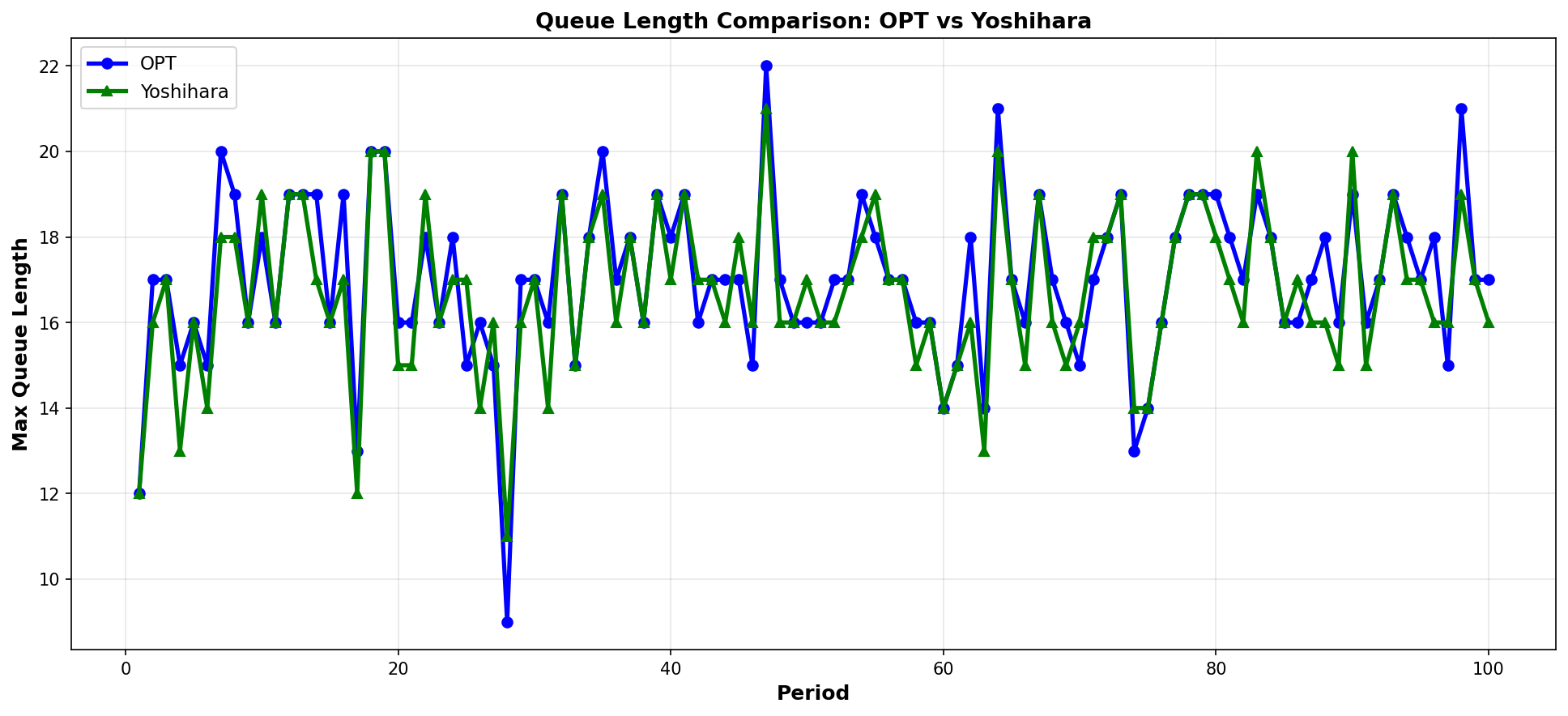}
\caption{Queue length: Yoshihara's vs optimization}
\label{fig:yoshi_queue}
\end{figure}

\begin{figure}[htbp]
\centering
\includegraphics[width=0.55\linewidth]{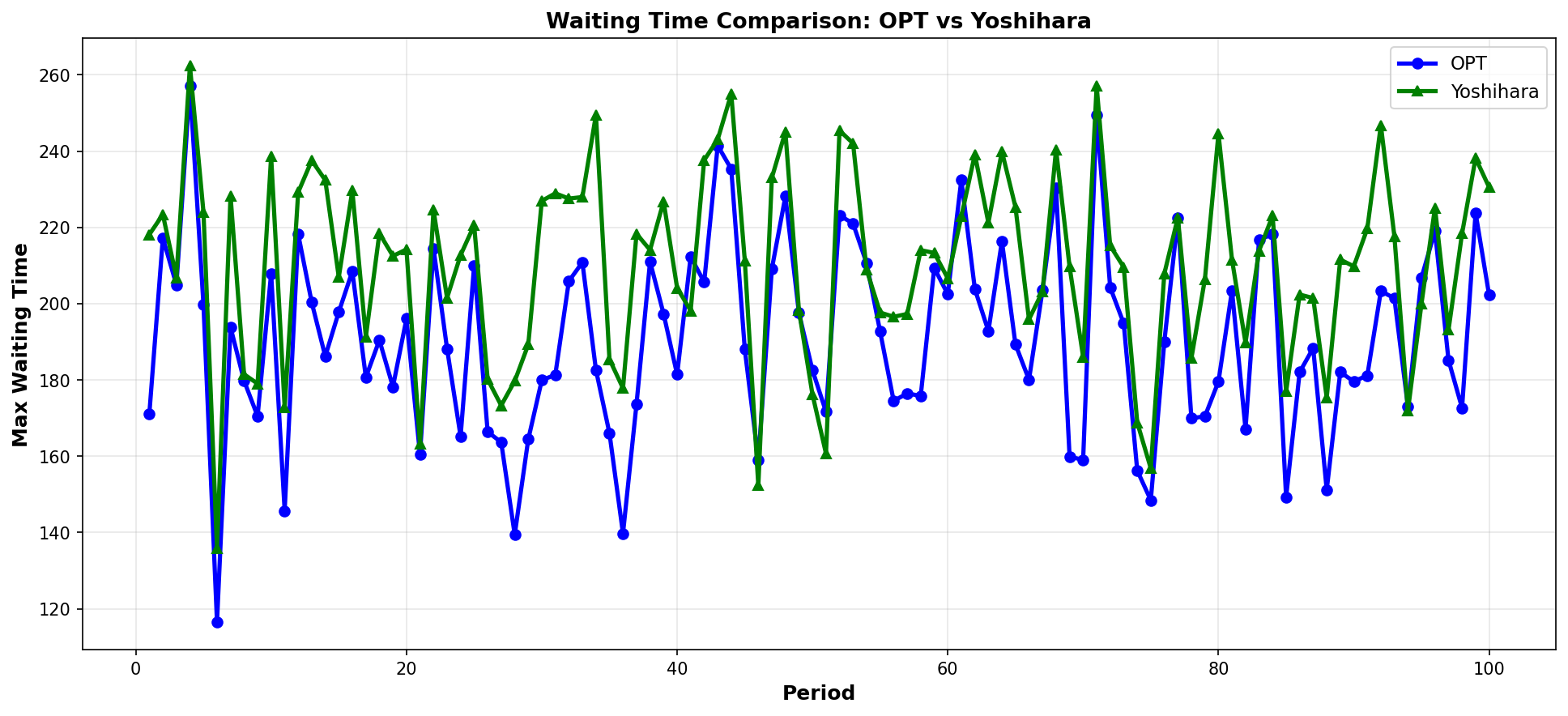}
\caption{Max delay: Yoshihara's vs optimization}
\label{fig:yoshi_wait}
\end{figure}

\begin{figure}[htbp]
\centering
\includegraphics[width=0.55\linewidth]{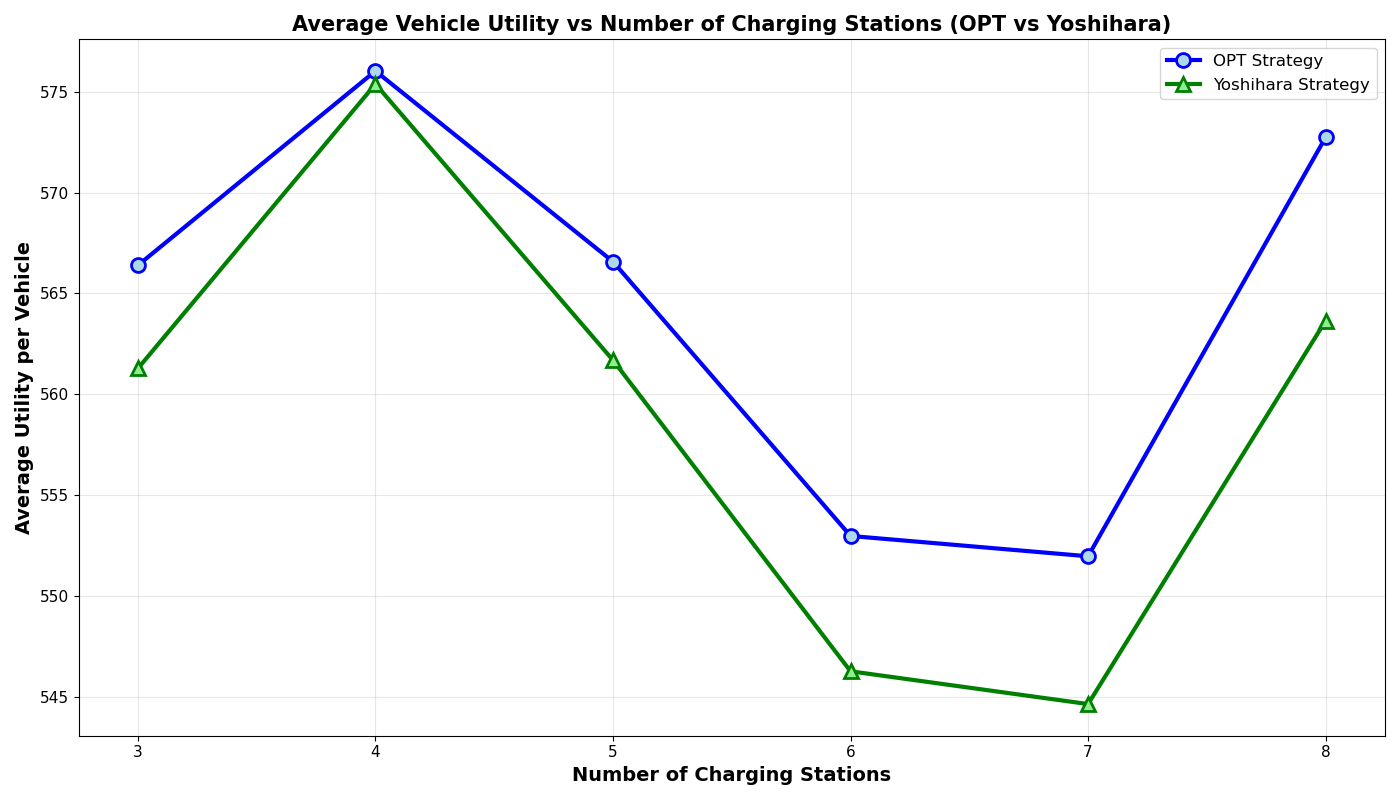}
\caption{Scaling: average utility vs number of stations}
\label{fig:yoshi_scale}
\end{figure}

We also compare our solution with the allocation algorithm of \citet{yoshihara_non-cooperative_2020} under the same environment configuration
as in the free-choice comparison. Figures~\ref{fig:yoshi_queue}--\ref{fig:yoshi_scale} summarize the comparison.

The key difference lies in Stage~2.
Our mechanism solves a welfare-maximizing capacitated assignment under Stage~1 quotas and feasibility constraints, without relying on station-side rejection.
By contrast, matching-based approaches may incorporate station preferences and rejection behavior, which can affect both allocation efficiency and congestion outcomes.
In particular, even when Stage~1-type congestion control is similar, differences in Stage~2 allocation logic can lead to different utility and delay profiles.

\section{Participation Decision Analysis}\label{sec:participation}
\paragraph{Purpose and link to Stage~1--2.}
Stages~1--2 characterize how a coordinated platform can (i) control congestion through station-level quotas and
(ii) improve allocative efficiency through a capacitated assignment.
In this subsection, we introduce a reduced-form participation model to capture the adoption threshold of such a platform.
The key parameters summarize the \emph{expected incremental gains from coordination} (as delivered by Stages~1--2),
the \emph{spillover benefits to nonparticipants}, and the \emph{coordination/overhead costs} of operating the platform.
\subsection{A reduced-form adoption model}
\paragraph{Players and participation rate.}
Consider a large population of EV users. Each user chooses whether to join the coordinated platform (J) or not join (NJ).
Let $m\in[0,1]$ denote the fraction of users who join (participation rate).

\paragraph{Payoffs (reduced form).}
Let $\pi^{J}(m)$ and $\pi^{NJ}(m)$ denote the expected utilities of a representative user who joins and does not join, respectively.
We model:
\begin{align}
\pi^{J}(m)  &= A m - B m^2 - c, \label{eq:payoff_join}\\
\pi^{NJ}(m) &= G m, \label{eq:payoff_notjoin}
\end{align}
where $A>0$ captures the strength of network benefits from coordination (e.g., better matching and congestion reduction),
$G\in[0,A)$ captures spillover benefits enjoyed by nonparticipants (free-riding),
$B>0$ captures convex coordination/overhead or congestion costs increasing with platform scale,
and $c\ge 0$ is the (per-user) cost of joining (e.g., registration, privacy, compliance, or fees).

\paragraph{Join condition and sustainable participation interval.}
A user joins if joining yields at least as much utility as not joining:
\begin{equation}\label{eq:join_condition}
\pi^{J}(m)\ge \pi^{NJ}(m)
\quad\Longleftrightarrow\quad
(A-G)m - Bm^2 - c \ge 0.
\end{equation}
Let $\Delta=(A-G)^2-4Bc$.
If $\Delta<0$, inequality \eqref{eq:join_condition} has no real interior solution.
If $\Delta\ge 0$, define the two roots
\begin{equation}\label{eq:m_thresholds}
m_{1,2}=\frac{(A-G)\mp\sqrt{\Delta}}{2B}.
\end{equation}
Then joining is individually rational for
\begin{equation}\label{eq:sustainable_interval}
m\in[m_1,m_2]\cap[0,1].
\end{equation}

\paragraph{Interpretation.}
For small $m$, network benefits are insufficient to offset the join cost $c$.
For large $m$, convex overhead/congestion costs $Bm^2$ dominate.
Thus, the platform is sustainable only for intermediate participation rates, consistent with threshold-type adoption and
the need for early-stage incentives (e.g., subsidies) to overcome the cold-start problem.

\subsection{Optional heterogeneity extension}
To reflect user heterogeneity while retaining tractability, one may allow heterogeneous join costs $c_n$.
Then user $n$ joins if $(A-G)m - Bm^2 \ge c_n$.
Let $F_C(\cdot)$ denote the CDF of $c_n$ in the population. The participation rate satisfies the fixed-point condition
\begin{equation}\label{eq:hetero_fixedpoint}
m = F_C\!\big((A-G)m - Bm^2\big),
\end{equation}
which admits standard threshold dynamics and can be used in numerical experiments if desired.
In the main text, we focus on the homogeneous-$c$ benchmark \eqref{eq:payoff_join}--\eqref{eq:payoff_notjoin}
to highlight the participation threshold mechanism.

\section{Discussion}\label{sec:discussion}
This paper provides a simulation-ready two-stage framework for EV charging allocation that targets worst-case congestion in Stage~1 and preserves welfare in Stage~2 under capacity constraints. The participation module provides a transparent characterization of adoption thresholds and free-riding incentives.

Several limitations remain. First, the queue-control layer relies on a tractable approximation for outflow and a quasi-stationary interpretation within epochs. Second, the reduced-form participation module summarizes Stage~1--2 gains using parameters $(A,G,B,c)$ rather than deriving them endogenously. Third, ``cold start'' in the first epoch is unavoidable without prior data; our proportional initialization follows the original simulation spirit.

Future work includes calibrating congestion and participation parameters using real charging-network data, introducing time-varying prices and reservation/slot mechanisms, and improving computational efficiency for large-scale deployments.

\section{Conclusion}\label{sec:conclusion}
We propose a two-stage optimization framework for EV charging allocation. Stage~1 determines station quotas to control extreme congestion, and Stage~2 solves a utility-maximizing capacitated assignment using closed-form charging amount decisions. We also present a reduced-form participation model that yields adoption thresholds under network benefits, spillovers, and coordination costs. Numerical experiments suggest substantial reductions in worst-case congestion while maintaining competitive average utility.
\bibliographystyle{apalike}
\bibliography{example}

@article{yoshihara_non-cooperative_2020,
	title = {Non-Cooperative Optimization Algorithm of Charging Scheduling for Electric Vehicle},
	volume = {13},
	issn = {1882-4889, 1884-9970},
	url = {https://www.tandfonline.com/doi/full/10.9746/jcmsi.13.265},
	doi = {10.9746/jcmsi.13.265},
	pages = {265--273},
	number = {6},
	journal = {{SICE} Journal of Control, Measurement, and System Integration},
	shortjournal = {{SICE} Journal of Control, Measurement, and System Integration},
	author = {Yoshihara, Miyu and Hafizulazwan Mohamad Nor, Mohamad and Kono, Akari and Namerikawa, Toru and Qu, Zhihua},
	urldate = {2025-06-24},
	year = {2020},
	langid = {english},
}

@article{2017Distributed,
  title={Distributed Scheduling and Cooperative Control for Charging of Electric Vehicles at Highway Service Stations},
  author={ Gusrialdi, Azwirman  and  Qu, Zhihua  and  Simaan, Marwan A. },
  journal={IEEE Transactions on Intelligent Transportation Systems},
  volume={18},
  number={10},
  pages={2713-2727},
  year={2017},
}

\end{document}